\documentclass[aps,preprint]{revtex4}
\usepackage[dvips]{graphicx}
\newcommand{\futoi}[1]{\mbox{\boldmath$#1$}}

\begin{document}

\title
{
Current-induced vortex displacement and annihilation \\ 
in a single Permalloy disk
}

\author
{ 
T. Ishida 
}
\affiliation{
Institute for Solid State Physics, University of Tokyo 
5-1-5 Kashiwanoha, Kashiwa, Chiba 277-8581, Japan
}

\author
{ 
T. Kimura 
}
\email{kimura@issp.u-tokyo.ac.jp}
\affiliation{
Institute for Solid State Physics, University of Tokyo 
5-1-5 Kashiwanoha, Kashiwa, Chiba 277-8581, Japan
}
\affiliation{
RIKEN FRS, 2-1 Hirosawa, Wako, Saitama 351-0198, Japan 
}
\affiliation{
CREST, JST, Honcho 4-1-8, Kawaguchi, Saitama, 332-0012, Japan 
}

\author
{
Y. Otani 
}
\affiliation{
Institute for Solid State Physics, University of Tokyo 
5-1-5 Kashiwanoha, Kashiwa, Chiba 277-8581, Japan
}
\affiliation{
RIKEN FRS, 2-1 Hirosawa, Wako, Saitama 351-0198, Japan 
}
\affiliation{
CREST, JST, Honcho 4-1-8, Kawaguchi, Saitama, 332-0012, Japan 
}

\date{\today}
\begin{abstract}
The induced motion of a magnetic vortex in a micron-sized ferromagnetic disk 
due to the DC current injection is studied by measuring planar Hall effect.  
The DC current injection is found to induce the spin torque 
that sweeps the vortex out of the disk at the critical current while bias magnetic field 
are applied.  
The current-induced vortex core displacement deduced from the change in planar Hall resistance 
is quantitatively consistent with theoretical prediction.  
Peak structures similar to those originated from spin wave excitations 
are observed in the differential planar Hall resistance curve.

\end{abstract}

\maketitle
Recently, controlling domain structures in the patterned magnetic structures 
by using electric currents has drawn much attention 
because of its potentiality for device application as well as novel spin-related physics.
\cite{Yamaguchi, Saitoh, Klaui}  
Electric currents flowing in ferromangets generate spin currents which 
transfer the transverse component of the spin angular momentum to the local 
magnetic moments while traveling through spatially 
varying magnetic structures such as domain walls.\cite{Berger, Slon}  
This spin torque, the magnitude of which is proportional to the spatial derivative 
of the magnetization $\nabla \futoi{M}$\cite{Baza, Nakatani}, 
drives a domain wall.  
For example, the magnetic force microscope observation revealed that 
both the transverse and vortex-like domain walls were driven by application of 
high DC currents.\cite{Yamaguchi}  
One of the important issues to be solved is to clarify experimentally 
how the vortex-like wall feels the spin torque.  
Since the magnetic vortex is a kind of a confined vortex-like wall, 
the vortex motion due to the spin torque can be expected.  
Moreover, a large spin torque will be exerted on a vortex core of 
the exchange length in the order of a few nanometers.\cite{Science}   
Very recently, the vortex motion due to the spin torque has been 
investigated theoretically.\cite{Shibata2}  
They predict that the spin torque induces the force normal to the applied current 
expressed as $\futoi{G} \times \futoi{v_s}$, 
where $\futoi{G}$ and $\futoi{v_s}$ are, respectively, 
the gyrovector, defined as the product of the vortex polarity and vorticity  
and the drift velocity of the electron spins.  
The vortex displacement is expected proportional to the spin current density.  
So far, no experimental studies have been reported concerning the current-induced motion 
of the vortex core confined in a magnetic disk.    
Here we experimentally investigate the influence of the spin torque 
on the vortex core in the magnetic disk.

A Permalloy (Py) disk 2 $\mu$m in diameter and 30 nm in thickness is fabricated 
by means of electron-beam lithography and lift-off techniques.  
The Py layer is evaporated by using an electron-beam gun at the base pressure of $10^{-10}$ Torr.   
In order to investigate the vortex displacement under DC current injection, 
four Cu probes 80 nm in thickness are connected to the Py disk.  
The inset of Fig.\ 1 is a scanning-electron-microscope (SEM) image of a fabricated device.  
The resistivity of the Py layer is 14.6 $\mu\Omega$cm  at room temperature.

We first study the magnetization process of a prepared Py disk by 
using high-sensitive micro Magneto-Optical Kerr Effect (MOKE) magnetometer 
laser beam of which diameter is less than 5 $\mu$m.\cite{Akahane}  
Figure 1(a) shows a Kerr hysteresis loop for a single Py disk.  
The Kerr loop is obtained by averaging measurements 200 times.  
When the magnetic field is decreased from the saturated state, 
the Kerr signal gradually decreases, followed by an abrupt jump at the nucleation field $H_{n}$, 
where a single magnetic vortex is formed in the Py disk. 
In the remanent state, the vortex stays at the disk center.  
When the external magnetic field is applied, 
the vortex is shifted perpendicular to the magnetic field 
to increase the net magnetization along the field. 
At the annihilation field $H_{an}$, the vortex vanishes and 
the magnetization state turns out to be a single domain.  
Thus, the magnetization reversal of the disk 
takes place accompanied by nucleation and annihilation of the single vortex.\cite{Otani2}  
The magnetization reversal of the disk is also examined 
by measuring the planar Hall resistance (PHR) given by $V_t/I$ with the probe configuration 
in the inset of Fig.\ 1(a), where $V_{t}$ is the transverse voltage 
induced in between voltage probes.  
Note that the AC excitation current of 5 $\mu$A is small enough to disregard 
its influence on the domain structure.\cite{Kimura}  
To avoid the influence of inhomogeneous current flow in the 
vicinities of Cu probes, where the current distribution is disturbed,
as shown in the inset of Fig.\ 1 (b), 
the magnetic field is applied at an angle $\phi = \pi/4$  
with respect to the average current direction for the PHR measurements.  
In this case, the vortex core moves along the diagonal direction $\phi = 3\pi/4$.  
Moreover, applying the magnetic field at $\phi = \pi/4$ 
produces large PHR signal 
because of the angular variation given by $\sin 2\varphi$, 
where $\varphi$ is the angle between the magnetization and the current.\cite{Jan}  
As shown in Fig.\ 1(b), the PHR curve exhibits clear two abrupt changes, 
corresponding to the nucleation and annihilation. 
Thus, the PHR measurements yield detailed information on the single vortex motion in the magnetic disk.

The vortex motion under the DC current injection is studied 
by measuring the differential PHR $dV_{t}/dI$ with variable DC current 
in the range from $-10$ mA ($-1.2 \times 10^{11}$ A/m$^2$)  to 
10 mA ($1.2 \times 10^{11}$ A/m$^2$) 
superimposed on the AC exciting current.  
Here, the DC current is increased stepwise by 0.02 mA. 
Figure 2(a) shows $dV_{t}/dI$ as a function of the DC current 
in the absence of bias magnetic fields.  
$dV_t/dI$ varies parabolically with DC current, 
but is slightly asymmetric with respect to the current.  
The DC current dependence of $dV_t/dI$ is found to be expressed by 
$ a_2 I^2  + a_1 I + a_0 $.  
Here, $a_2 = 2.90 \times 10^{-5}$, $a_1 =-8.98 \times 10^{-5}$ 
and $a_0 = 2.52 \times 10^{-2}$.  
The parabolic dependence of $dV_{t}/dI$ 
on the DC current is attributable to Joule heating.\cite{Katine}  
By considering the fact that the spin torque gives rise 
to the linear dependence of the vortex core displacement on the current $I$,
the parabolic component is subtracted from the differencial PHR curve 
to obtain a blue line in Fig.\ 2(a).  
This linear current dependence seems consistent with the theoretical prediction 
if the displacement is proportional to the magnitude of PHR.  
Now, we estimate the vortex displacement from the changes in the PHR curve.  
In low magnetic fields, it is known that the vortex displacement is proportional 
to the applied external magnetic field.\cite{Otani2}  
The reversible linear change in the Kerr signal in the low magnetic fields 
in Fig.\ 1(a) corresponds to the vortex displacement.  
Therefore, the relation between the vortex displacement 
and the magnetic field is deduced from the slope of the Kerr magnetization curve.   
According to the theoretical study, the vortex moves perpendicular to the current.\cite{Shibata2}  
A similar vortex displacement is induced by applying the magnetic field along 
the average current direction corresponding to $\phi = 0$.  
We also measure the PHR at the magnetic field at $\phi = 0$.  
Although the PHR does not show the simple change near the nucleation and annihilation fields 
because of the disturbance of the Cu probes, 
the PHR shows the continuous change in the low magnetic field corresponding to the vortex displacement.    
Therefore, by comparing the Kerr loop and the PHR curve for $\phi =$ 0 measured in low magnetic fields, 
we can estimate the vortex displacement as a function of the PHR.  
In this way, the vortex displacement due to the DC current injection is deduced 
from the change in PHR.  
The experimentally obtained relation is $\delta$ [nm] $= 1.23 \times 10^{-10} J$[A/cm$^2$], 
where $J$ is the density of the DC current.  
This is quantitatively good agreement with the relation 
$\delta$ [nm] $= 0.76 \times 10^{-10} J$[A/cm$^2$], 
which is theoretically calculated.\cite{Shibata2}

In the above discussion, we assumed that the linear dependence of the change in PHR 
is caused by the vortex displacement due to the spin torque with 
no direct experimental evidence of the vortex motion.  
To verify that the DC current moves the vortex core, 
we examine the current-induced vortex annihilation.  
As mentioned previously, 
the vortex displacement due to the DC current injection is expected about ten nanometers 
even at the current of 10 mA.  
It thus seems difficult to annihilate the vortex only by the DC current injection 
without applying bias magnetic fields.  
Therefore, we measure the differential PHR under the DC current injection 
with applying a fixed bias magnetic field along $\phi = \pi/4$ .  
Before sweeping the current, the magnetic field has been scanned from $-1000$ Oe 
to set a desired value.  
Figure 2(b) shows the differential PHR as a function of the DC current 
at a fixed magnetic field of 160 Oe.  
The differential PHR curve shows large asymmetry with respect to the DC current.  
This asymmetry is more pronounced with increasing the bias field.  
This can be understood as follows.  
The vortex core moves perpendicular to the magnetic field, 
toward the edge of the disk with the applied magnetic field  
When the magnetic field is applied at $\phi=\pi/4$, 
the vortex core shifts along the diagonal direction corresponding to $\phi = 3\pi/4$.  
The numerical calculation of the current distribution inside the disk in Fig.\ 1(b) 
shows that 
the current near the edge flows in the azimuthal direction.  
Therefore, the direction of the vortex displacement due to the spin torque 
is the same as that due to the magnetic field near the annihilation field.  
According to the PHR curve for $\phi = \pi/4$ shown in Fig.\ 1(b), 
increasing or decreasing the PHR corresponds to the vortex displacement 
toward the edge or the center of the disk, respectively.  
Therefore, the negative current injection should induce the vortex annihilation 
when the positive magnetic field at $\phi =$ 45 deg is applied.   
The current induced vortex annihilation is observed in the magnetic field range 
from 180 Oe to 195 Oe.  
Figure 3(a) shows a typical differential PHR as a function of the DC current 
exhibiting the vortex annihilation.  
Once the vortex is swept out of the disk by the DC current injection, 
the vortex does not nucleate in the disk even if the DC current decreases.   
Figure 3(b) shows the differential PHR dependence on the DC current for various bias fields.  
The critical current $I_{an}$, where the vortex annihilates, 
decreases monotonically with increasing the bias field.  
This tendency is consistent with our expectation that 
the negative DC current exerts the torque which sweeps the vortex out of the disk.

Finally, we discuss the peak structures observed in the PHR in Figs.\ 3(a) and 3(b).  
The peak structures are observed right before the vortex annihilation 
in the bias magnetic field range from 170 Oe to 185 Oe 
and is observed only for the negative DC current.  
Figure 4(a) shows the differential PHR as a function of the current for various bias fields.  
The current $I_{P}$, where the abrupt change appears, 
monotonically decreases with increasing the bias field.  
This tendency is the same as that of $I_{an}$ 
and implies that the peak in the PHR 
is related to the magnetic domain structure.  
Similar current-induced resistance peaks have been observed in vertical structures 
consisting of a point contact between a single ferromagnetic layer and nonmagnetic probe\cite{Chien} 
and a nanopillar consisting of the magnetic multilayer\cite{Ralph} 
which are explained as the spin-dependent scattering due to the spin precession.  
The peak structures in the PHR observed in the present experiment 
should also be caused by the same mechanism.  
However, in the present experiment, 
the position of the resistance peak decreases with increasing the bias field.  
This tendency is opposite to that observed in the vertical structures.\cite{Chien}  
In the vertical structures, the peak corresponding to the spin precession 
is observed when the uniform magnetization state is stabilized by a strong magnetic field.  
The spin precession is induced when the spin torque balances the damping torque.  
Since the damping torque is proportional to the applied magnetic field, 
the critical current for the spin precession increases with increasing the applied field.\cite{Shibata3}  
However, in the present experiment, 
the vortex core is pushed to the edge of the disk, 
leading to an increase in spin torque because of the increment of $\nabla \futoi{M}$.  
Therefore, the critical current $I_{p}$ decreases with increasing the bias field.  
As shown in Fig.\ 4(b), the abrupt HR change does not show the hysteresis behavior 
and it is also consistent with the interpretation base on the spin precession due to the spin torque.

\hspace*{0.5cm}
In conclusion, we study the responce of the magnetic vortex under the DC current injection by using PHE.   
The observed change in the PHE due to the DC current injection is 
consistent with the theoretical perdition that 
the vortex displacement is proportional to the current.  
The current-induced vortex annihilation is observed 
when a fixed bias magnetic field close to the annihilation field is applied.  
The peak structures are observed in the differential PHR curve right before 
the current-induced vortex annihilation, which may be caused by 
by the spin torque similar to those observed in the vertical magnetic structures.

\section*{Acknowledgement}
The authors would like to thank Y. Iye and S. Katsumoto 
for the use of micro-fabrication facilities 
and J. Shibata and Y. Nakatani for valuable discussions.

\newpage

\begin{figure}
\caption
{
(a) Micro MOKE hysteresis loop for a single Py disk measured at RT.  
The inset shows a Scanning-electron-microscope image of the fabricated Py disk.  
(b) Planar Hall resistance curve of the Py disk measured at RT.  
The inset shows the calculated spatial distribution of the electrical potential 
in the Py disk with the probe configuration for the PHE.   
}

\caption
{
(a) Differential planar Hall resistance in the absence of the magnetic field 
as a function of the DC current (black line) 
and that except for the parabolic component (blue line).  
(c) Differential planar Hall resistance at the magnetic field of 160 Oe 
as a function of the DC current.  
}

\caption
{
(a) Differential planar Hall resistance as a function of the 
DC current at the magnetic field of 190 Oe
(b) Differential planar Hall resistance as a function of the DC current 
and the bias magnetic field in the range from $178$ Oe to $195$ Oe.  
The DC current is swept from 0 to $-10$ mA.  
}

\caption
{
(a) Differential planar Hall resistance as a function 
of the DC current for various the bias magnetic fields 
in the range from $170$ Oe to $188$ Oe.  
The differential PHR is measured with sweeping the DC current 
from 0 to $-10$ mA in a fixed applied magnetic field.  
(b) Differential Hall resistance as a function of the 
DC current at the magnetic field of 170 Oe
}

\end{figure}

\vspace*{1cm}
\newpage

\newpage
\vspace*{2cm}
\begin{center}
\includegraphics[scale=0.5]{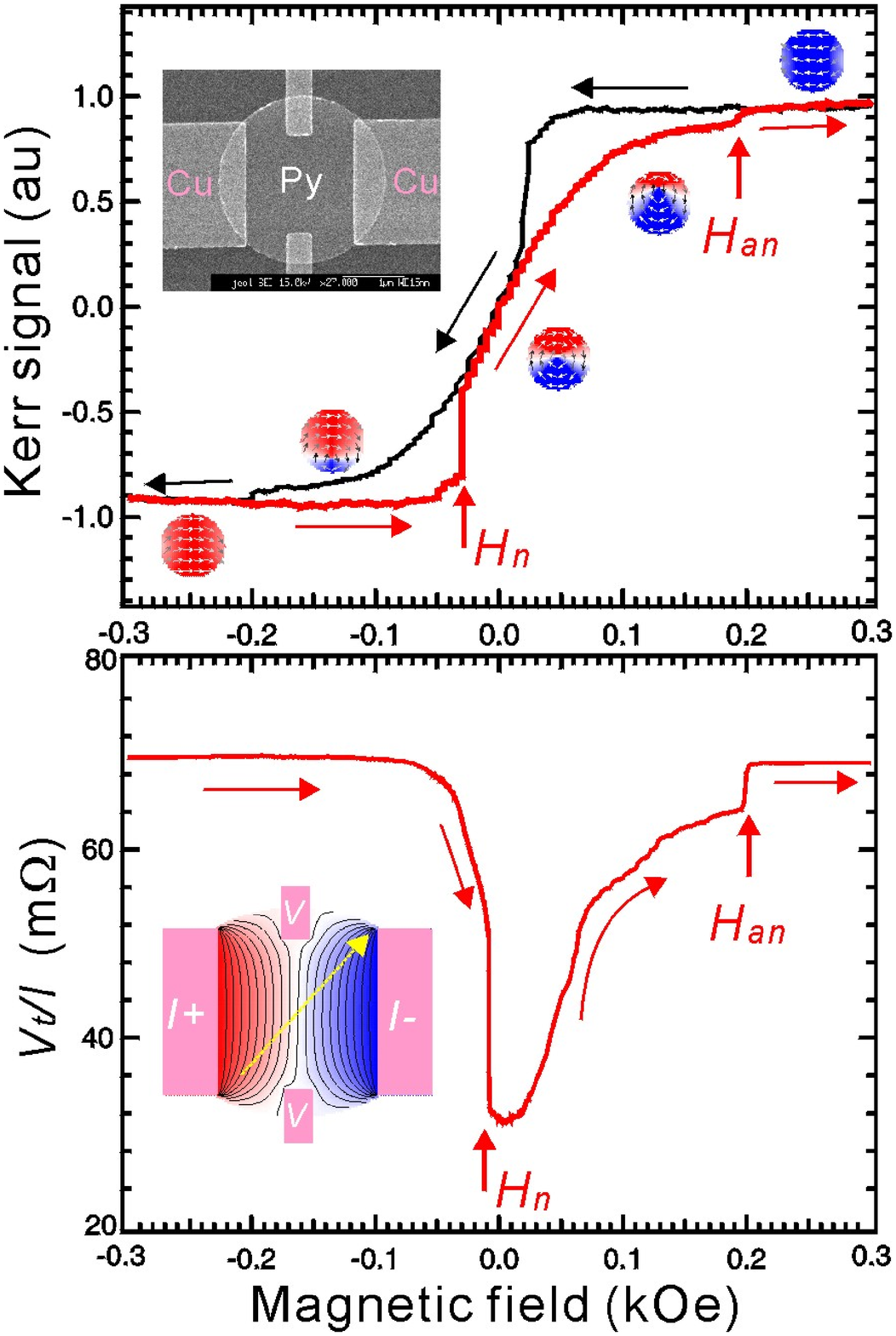}
\end{center}
\vspace*{1cm}
\begin{center}
Fig.\ 1 Kimura et al. \\ 
Color only in on-line
\end{center}

\newpage
\begin{center}
\vspace*{3cm}
\includegraphics[scale=0.5]{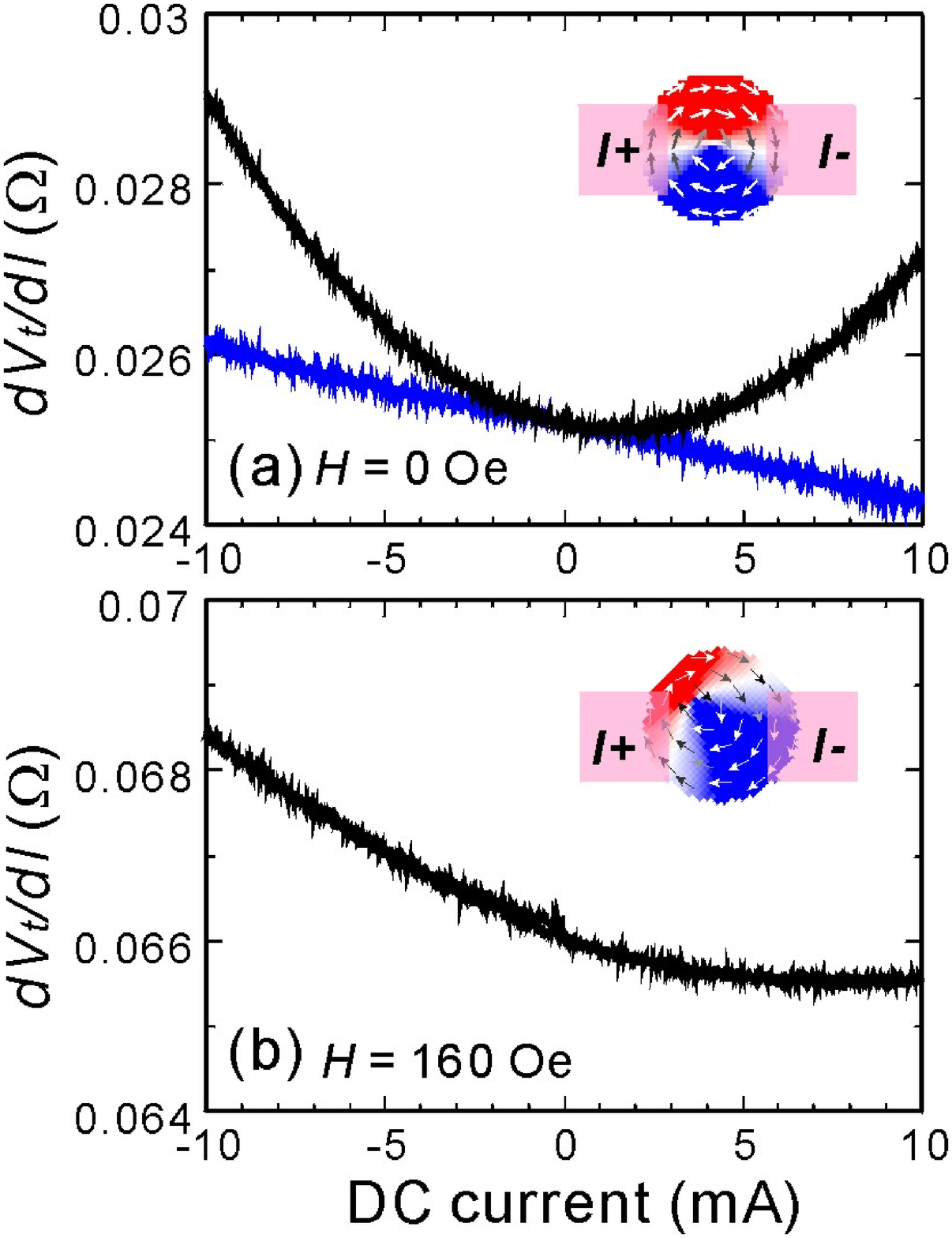}
\end{center}
\vspace*{1cm}
\begin{center}
Fig.\ 2 Kimura et al. \\
Color only in on-line
\end{center}

\newpage
\begin{center}
\vspace*{3cm}
\includegraphics[scale=0.5]{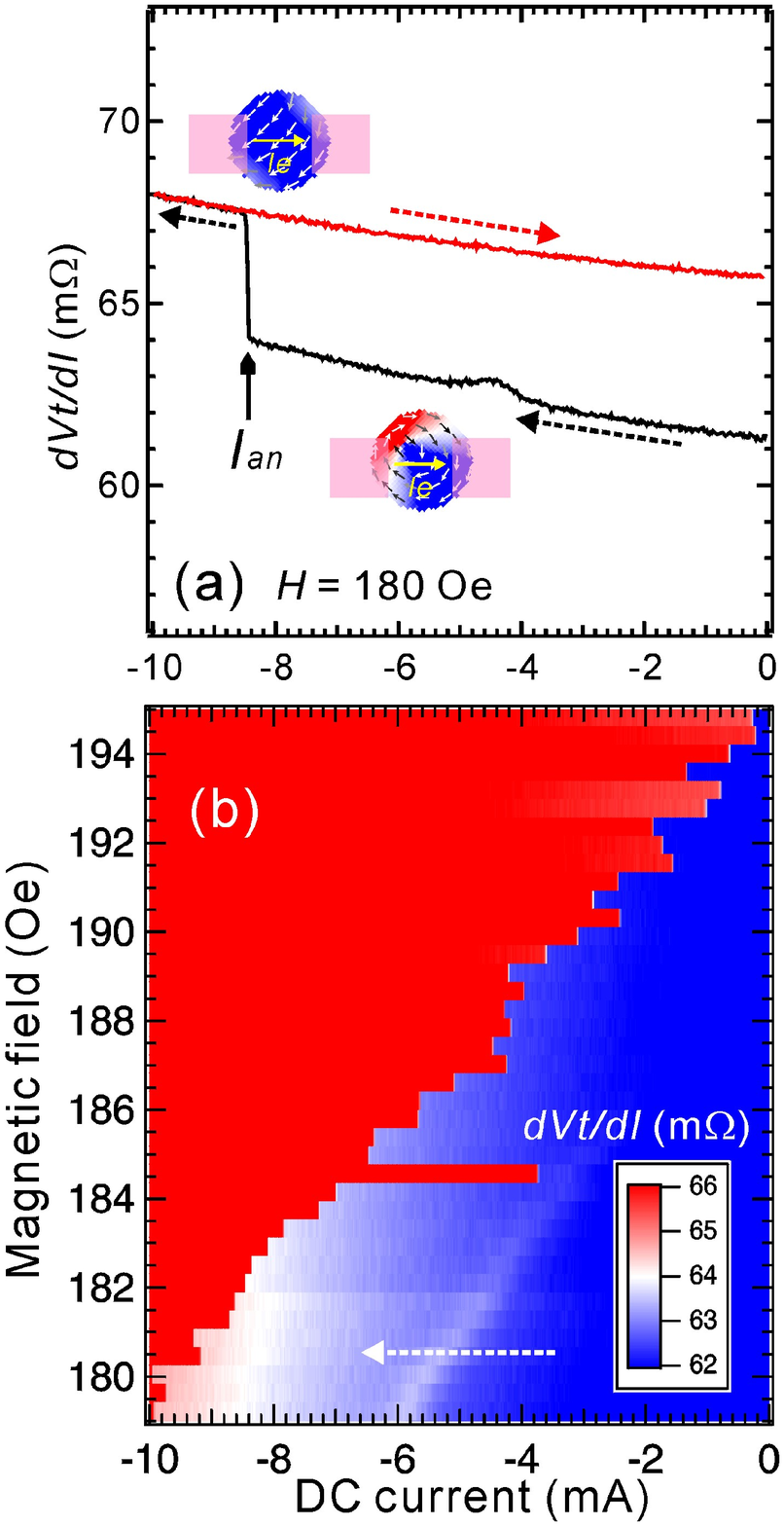}
\end{center}
\vspace*{1cm}
\begin{center}
Fig.\ 3 Kimura et al. \\
Color only in on-line
\end{center}

\newpage
\vspace*{2cm}
\begin{center}
\includegraphics[scale=0.5]{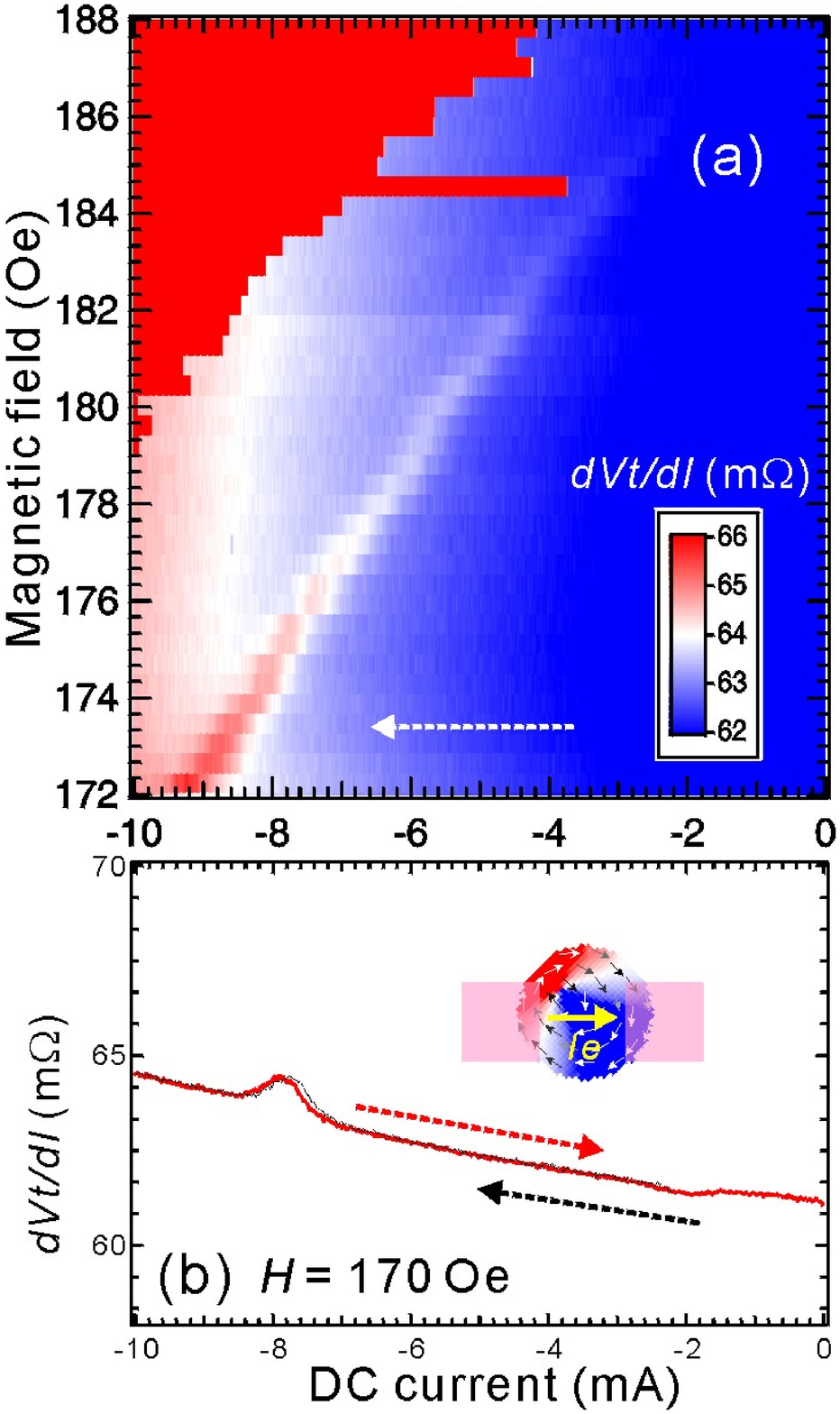}
\end{center}
\vspace*{1cm}
\begin{center}
Fig.\ 4 Kimura et al.
Color only in on-line
\end{center}


\newpage
\begin{references}


















\bibitem{Yamaguchi}
A. Yamaguchi, T. Ono, S. Nasu, K. Miyake, K. Mibu, and T. Shinjo, Phys. Rev. Lett. 92, 077205 (2004).

\bibitem{Saitoh}
E. Saitoh, H. Miyajima, T. Yamaoka, and G. Tatara 
Nature 432, 203 (2004)


\bibitem{Klaui}
M. Klaui, C. A. F. Vaz, J. A. C. Balnd, W. Wernsdorfer, G. Faini, E. Cambril, and L. J. Heyderman, 
Appl. Phys. Lett. 83, 108 (2003)


\bibitem{Berger}
L. Berger, J. Appl. Phys. 55, 1954 (1984), 
L. Berger, J. Appl. Phys. 71, 2721 (1992).


\bibitem{Slon}
J. C. Slonczewski, J. Magn. Magn. Mater. {\bf 159} (1996) L1


\bibitem{Baza}
Ya. B. Bazaliy, B. A. Jones, and Shou-Cheng Zhang 
Phys. Rev. B 57, R3213 (1998)


\bibitem{Nakatani}
A. Thiaville, Y. Nakatani, J. Miltat, and N. Vernier 
J. Appl. Phys. 95, 7049 (2004) 


\bibitem{Science}
A. Wachowiak, J. Wiebe, M. Bode, O. Pietzsch, M. Morgenstern, and R. Wiesendanger, 
Science 298, 577 (2002). 



\bibitem{Shibata2}
J. Shibata, Y. Nakatani, G. Tatara, H. Kohno and Y. Otani, 
cond-mat/0508599





\bibitem{Akahane}
K. Akahane, T. Kimura, and Y. Otani, 
J. Mag. Soc.Jpn., 28, 122 (2004)


\bibitem{Otani2}
K. Yu. Guslienko, V. Novosad, Y. Otani, H. Shima, and K. Fukamichi,  
Phys. Rev. B 65, 024414 (2002)


\bibitem{Kimura}
T. Kimura, Y. Otani, and J. Hamrle 
Appl. Phys. Lett. 87, 172506 (2005) 


\bibitem{Jan}  
J. P. Jan, Solid State Phys. {\bf 5} (1957) 17.


\bibitem{Katine}
J. A. Katine, F. J. Albert, and R. A. Buhrman 
Appl. Phys. Lett. 76, 354 (2000) 


\bibitem{Chien}
Y. Ji, C. L. Chien, and M. D. Stiles, Phys. Rev. Lett. 90, 106601 (2003).

\bibitem{Ralph}
I. Kiselev, J. C. Sankey, I. N. Krivorotov, N. C. Emley, R. J. Schoelkopf, R. A. Buhrman, D. C. Ralph
Nature 425, 380-383 (2003)


\bibitem{Shibata3}
J. Shibata, G. Tatara, and H. Kohno 
Phys. Rev. Lett. 94, 076601 (2005)




\end{references}
\end{document}